\documentclass[a4paper]{jpconf}
\usepackage{amsmath,mathtools}
\usepackage{amsfonts} 
\usepackage{graphicx} 
\usepackage{xcolor}
\usepackage{physics}
\usepackage[final]{hyperref}

\begin{document}
\title{Classical and relavistic simple wave problems solved with Smoothed Particle Hydrodynamics}

\author{J V O Caetano$^1$, L S Nowacki$^1$, V S Fran\c{c}\~ao$^1$, R Hirayama$^2$,
K~P~Pala$^1$, J~O~Sola$^1$ and F Grassi$^1$}
\address{$^1$ Instituto de F\'isica-Universidade de S\~ao Paulo, Rua do Matão
Nr.1371 CEP 05508-090 S\~ao Paulo-SP, Brazil}
\address{$^2$ Frankfurt Institute for Advanced Studies, Ruth-Moufang-Strasse 1, 60438 Frankfurt am Main, Germany}
\ead{jcaetano@usp.br,
nowacki.leandros@usp.br}

\begin{abstract}
To simulate the expansion of the matter created in relativistic nuclear collisions, codes in 3+1 dimensions are used and we are developing a new one. To benchmark such codes, the Sod's shock tube is often used. A closely related problem is the one-dimensional expansion of a gas into vacuum. In this paper, we study this problem classically and relativistically with the Smoothed Particle Method and test various techniques to improve the precision and speed of the solution.

\end{abstract}

\section{Introduction}
The matter created after a high energy collision, at the Relativistic Heavy-Ion Collider at Brookhaven National Laboratory or the Large Hadron Collider at CERN, can be treated as a fluid.

To solve the fluid mechanics equations, various  methods can be used, among them grid-based methods, with the mesh either  fixed in space (Eulerian approch) or 
fixed to the material (Lagrangian approach), or mesh-free methods such as the  Smoothed Particle Method,
or SPH for short, which we use here.
SPH was originally developed for 
studies in astrophysics by Lucy \cite{lucy} and  Gingold and Monaghan \cite{gingold} in 1977.
Later the method was extended to all sorts of problems \cite{monaghanreview,pricereview}.
Today the community of users,  SPHERIC\footnote{\url{<http://spheric-sph.org/>}},
 gathers researchers and industrial users and  created the Joe Monaghan prize in 2015.
 The method now enjoys great popularity, being used in video games,
 special effects in movies and virtual reality\footnote{ Gollum's fall into lava in the Lord of the Rings (2003) and several
  scenes in Superman Returns (2006) used SPH and were developed respectively
  by the companies
  Next Limit (who got a Technical ``Oscar'' for their software) and Tweak.}.
SPH is a Lagrangian method: the fluid is divided into  imaginary fluid particles, called SPH particles, and their motion is followed.

Once a code to solve the fluid mechanics equations is written, it must be benchmarked against known solutions. One example of such solutions is the Sod's shock tube, an infinitely long one dimensional tube filled with a perfect fluid, separated in a high density and a low density regions, by a membrane, removed instaneously at the initial time. This problem can be solved analytically in both classical and relativistic frameworks. A related problem is the case where, instead of low density matter, there is  vacuum. This  is the so-called simple wave problem. It too can be solved analytically both  classically and relativistically.  In high energy collisions, since matter may expand into vacuum, the simple wave problem should be an interesting test.
In this paper, we examine various ways to solve it numerically with the SPH method.


\section{Classical simple wave problem}

The problem consists of a long tube with  a membrane (at $z=0$). On the left of the membrane there is a gas and on the right, vacuum. 
At time $t=0$, the membrane is ruptured and the gas expands into the right half of the tube.
An analytical solution to the continuity and Euler equations can be found  \cite{simpleclass,simpleclassR}
for a polytropic equation of state $p=K\rho^{1+\frac{1}{n}}
\equiv K \rho^\gamma$.

For $z<-c_0 t$, the gas is unperturbed
while for $z>2c_0/(\gamma-1)\, t$, there is vacuum. Between these two values
\begin{equation}
  v(z,t)=\frac{2}{\gamma+1}\left(  c_0  + \frac{z}{t}\right).
  \label{eq:vsimple}
  \end{equation}
and
\begin{equation}
  \rho(z,t)=
\rho_0\left[ \frac{1}{c_0}\left(\frac{2 c_0}{\gamma+1}-\frac{\gamma-1}{\gamma+1}\frac{z}{t}\right) \right]^{\frac{2}{\gamma-1}}.
  \label{eq:rhosimple}
  \end{equation}
The index $_0$ indicates the initial unperturbed gas state (so $K=p_0\rho_0^{-\gamma}$,  $c_s^2=c_0^2(\rho/\rho_0)^{\gamma-1}$, etc.)

Equations \eqref{eq:rhosimple} and \eqref{eq:vsimple} take on some particular values that
are  useful to note for numerical checks
\begin{itemize}
\item  For $z=-c_0\,t$:
$\rho=\rho_0$ and $v=0$.
\item  For $z=2c_0/(\gamma-1)\, t$: $\rho=0$ and $v= 2c_0/(\gamma-1)$.
 \item  For
   $z=0$, whatever $t$: $\rho=\rho_0\left( \frac{2 c_0}{\gamma+1}\right)^{\frac{2}{\gamma-1}}$
   and   
$v=2c_0/(\gamma+1)$.
\end{itemize}

In the SPH approach \cite{monaghanreview,pricereview}, in one dimension, the mass density 
at some position $z$ is  approximated as
 \begin{equation}
   \rho(z)=\sum_{j=1}^N m_j W(z-z_j,h)
   \label{eq:sphrho}
 \end{equation}
 where we assumed the fluid divided into $N$  particles.
Conservation of the total total mass  $\int \rho (z) dz=\sum_{j=1}^{N} m_j$ is insured by requiring that
$W$  obeys the normalization condition $\int W(z-z',h) dz=1$.
Here we use for $W$ a cubic spline.
The SPH density summation \eqref{eq:sphrho}
satisfies the continuity equation  \cite{pricereview}.

The SPH equation of motion for a perfect fluid is
\begin{equation}
  \left(\frac{ dv_z}{dt}\right)_i=- \sum_{j=1}^N m_j \left(\frac{p_j}{\rho_j^2}
    + \frac{p_i}{\rho_i^2} \right)
\frac{d}{dz_i} W(z_i-z_j,h).
\label{eq:eulersph}
\end{equation}
The SPH expression (\ref{eq:eulersph}) leads
to conservation of total linear and angular momenta \cite{pricereview}.

\begin{figure}[h!]
\centering
\includegraphics[width=0.46\linewidth]{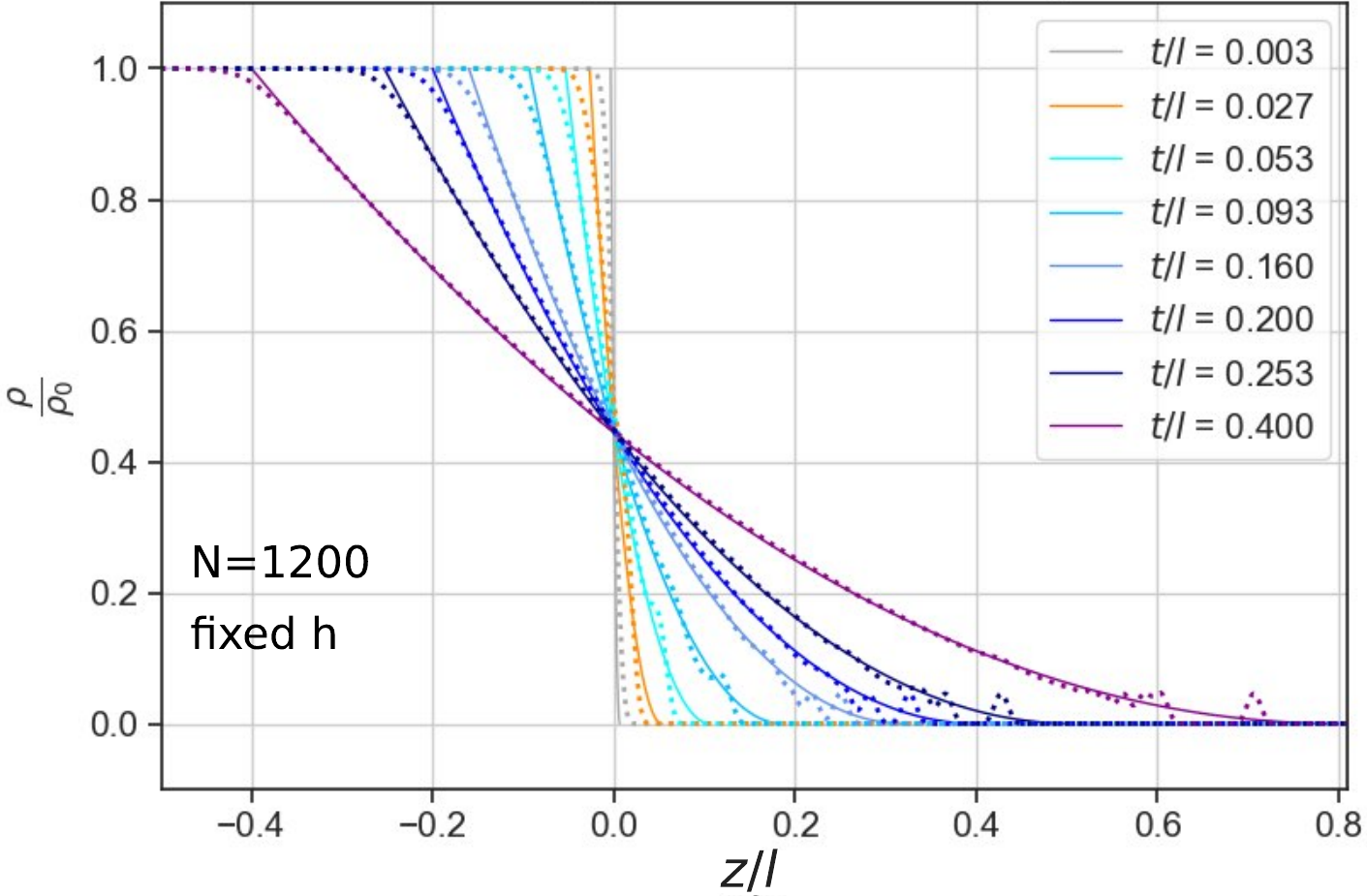}
  \includegraphics[width=0.46\linewidth]{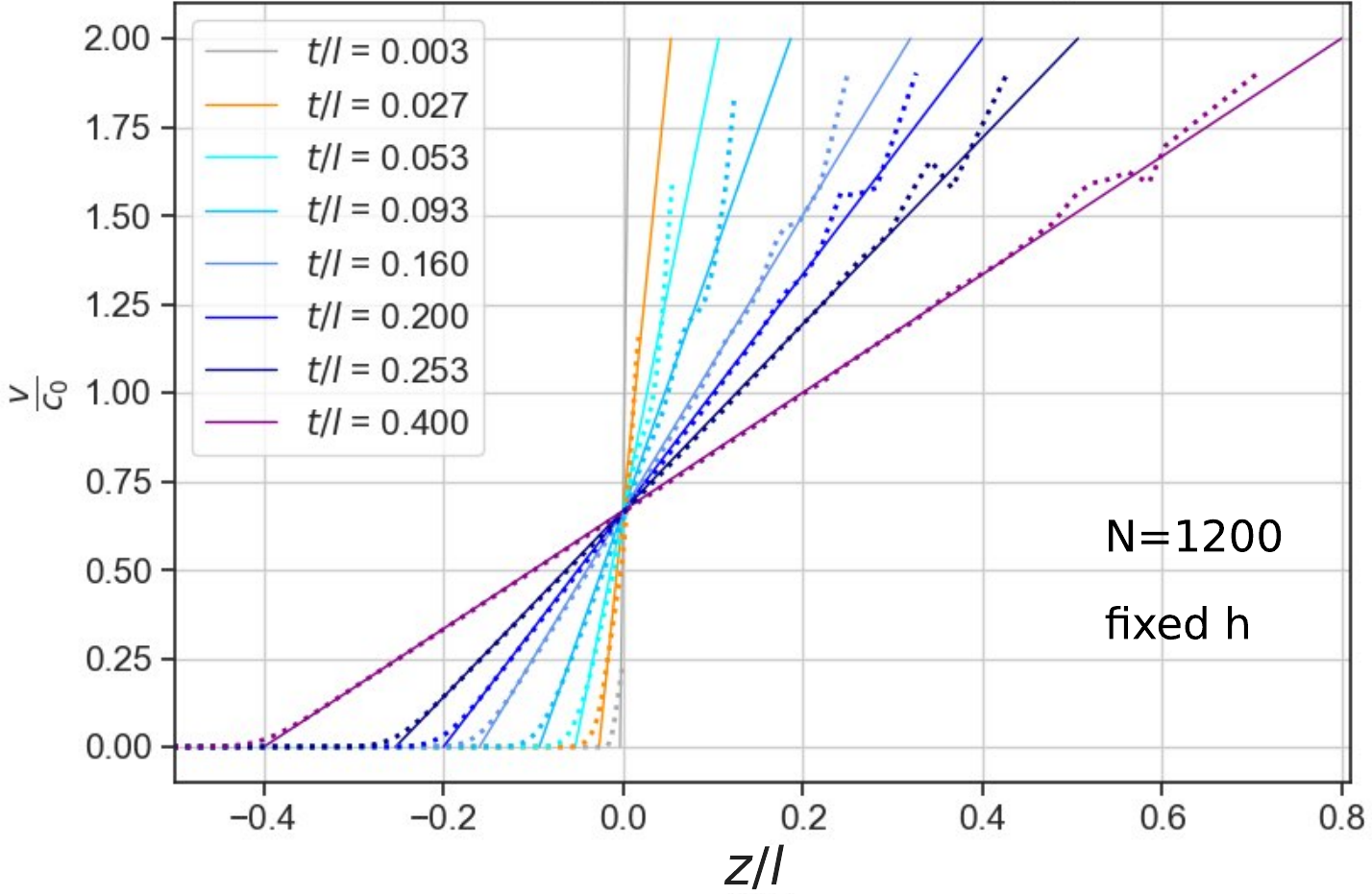}
  
\includegraphics[width=0.46\linewidth]{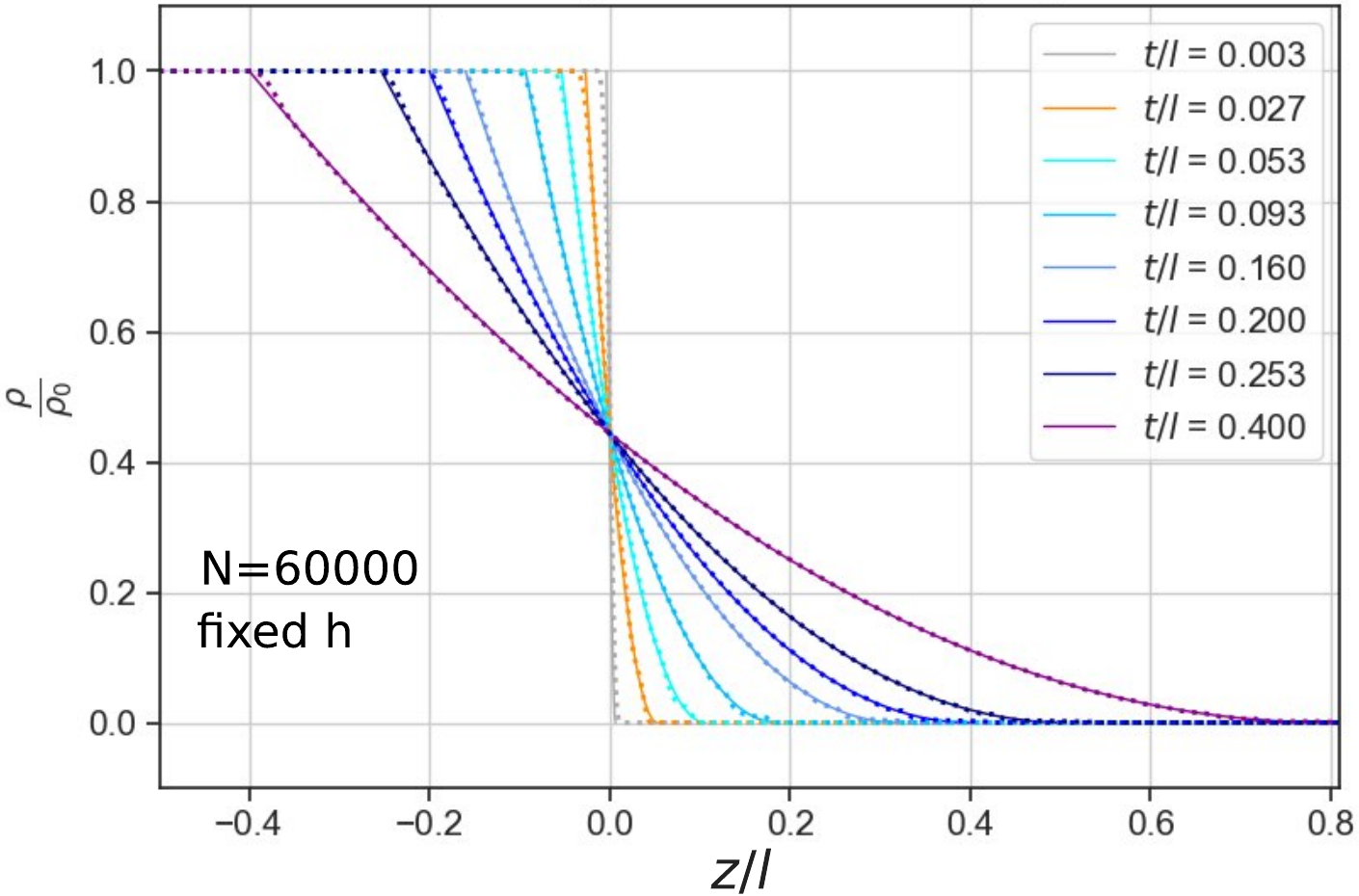}
  \includegraphics[width=0.46\linewidth]{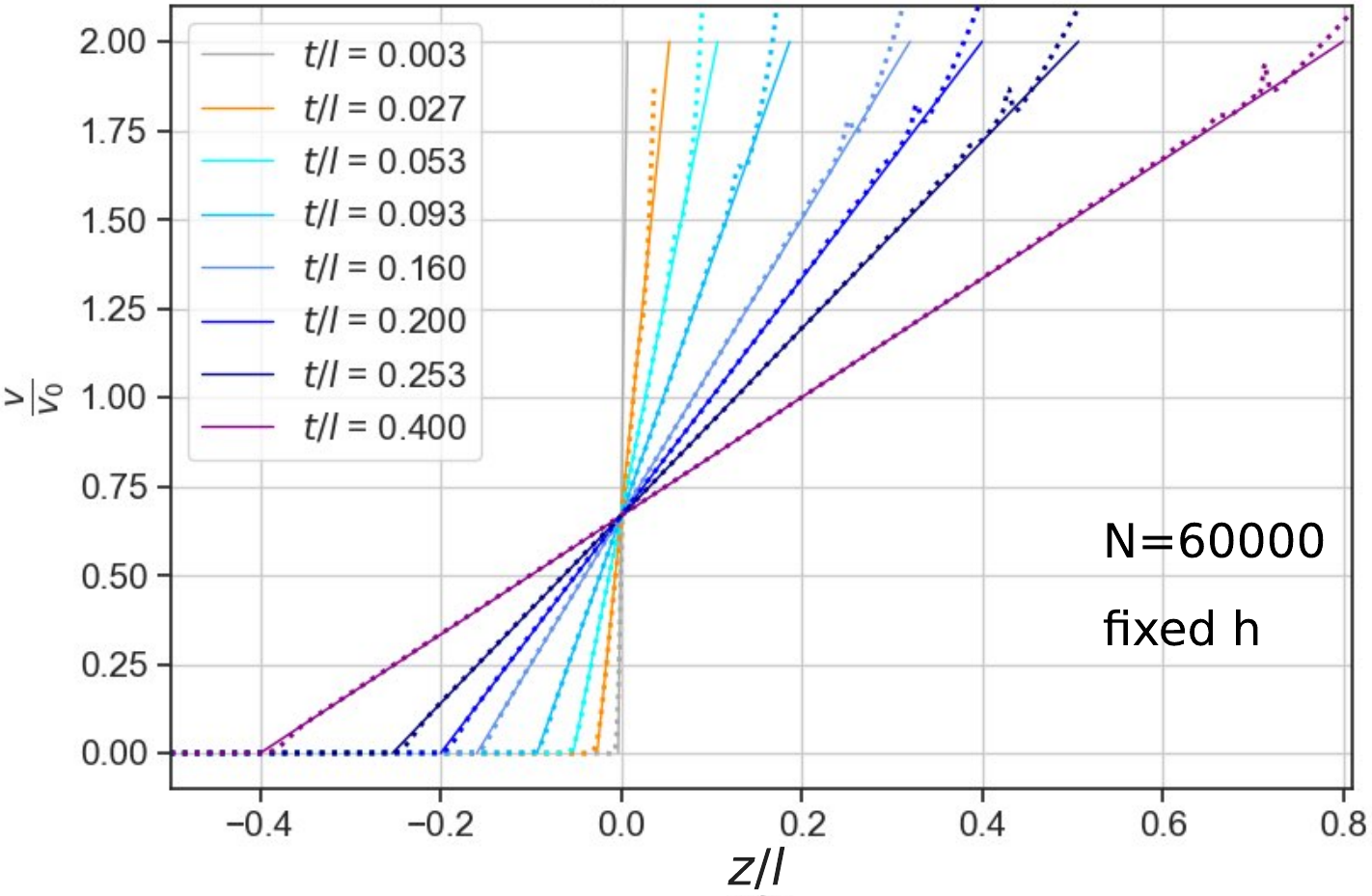}
  
\includegraphics[width=0.46\linewidth]{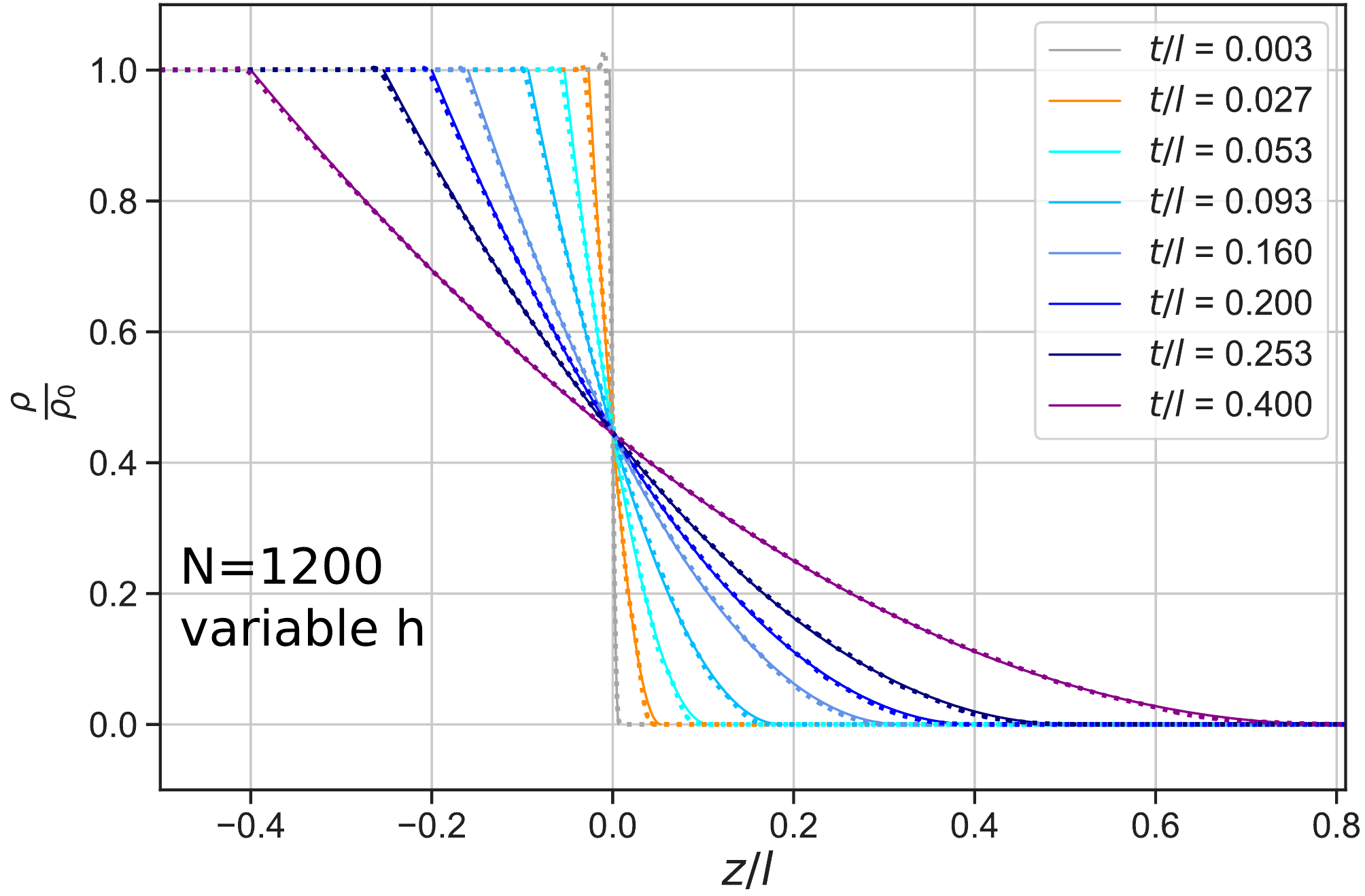}
  \includegraphics[width=0.46\linewidth]{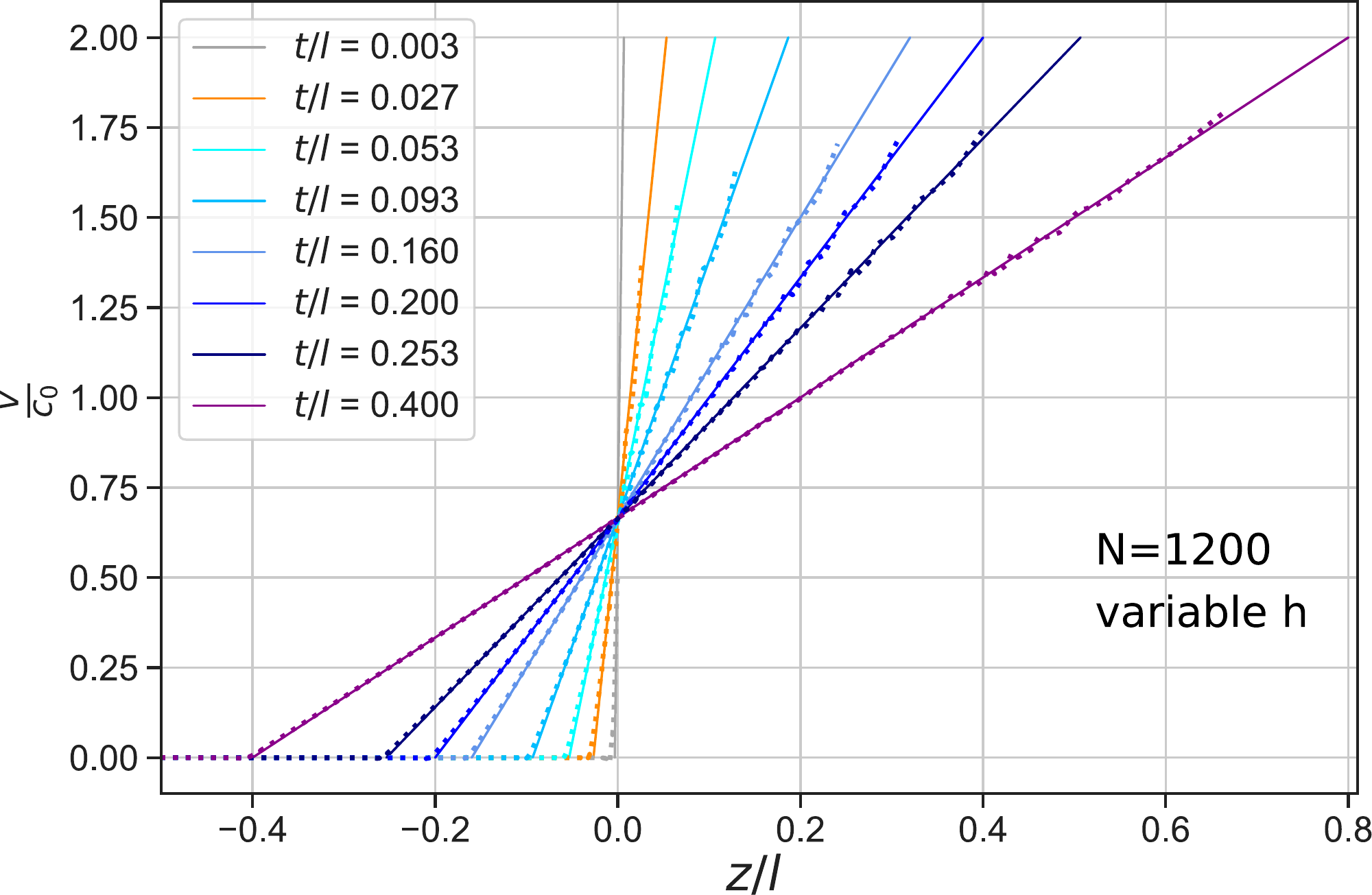}
  \caption{Solutions for
    one-dimensional gas expansion in vacuum: mass density (left) and fluid
    velocity (right). Solid lines represent the exact solution, dotted lines display the SPH evolution in three different configurations: fixed smoothing length $h$ and ``small'' number of interpolating points $N$ (upper), fixed $h$ and very large $N$ (middle), small $N$ and variable $h$ (lower).}
\label{fig:1dtube}
\end{figure}

This problem is not easy to solve precisely numerically because it involves discontinuities, in particular the velocity increases from 0 at $z\geq -c_0\,t$ to
$2c_0/(\gamma-1)$ at $z=2c_0/(\gamma-1)\, t$ and immediately falls back to 0 in the vacuum region.
Various solutions are shown in fig. \ref{fig:1dtube}: first with fixed smoothing length $h$ and two values of
the  number $N$ of SPH particles. Large $N$ allows for a better solution but  is time-consuming in higher dimensions, since the number of operations in a naive\footnote{For simulations with large $N$, we make use of a grid \cite{grid} as a neighbour search algorithm to reduce the number of operations.} SPH time step is $\sim\mathcal{O}(N^2)$.
Another possibility is using 
a variable $h$, related to the interparticle distance \cite{LiuLiu}. This allows for a better treatment of more diluted fluid regions without a large $N$.
The solution for variable $h$ is precise but the velocity did not reach the maximum value, so there is space for improvement.
Other possibilities that we investigated not (yet) leading to a great improvement are the inclusion of artificial viscosity and the use of higher order splines \cite{pricereview}.
We also plan to investigate the variable $h$ case with a different method \cite{pricereview}, particle splitting and imaginary particles.

\section{Relativistic simple wave problem}

The equations of conservation for energy and momentum
  $ \partial_\mu T^{\mu \nu}=0$ admit an exact solution \cite{simpleclass,Rischke}.
For $z<-c_0 t$, the fluid is unperturbed
 while
 for $z>t$, there is vacuum.  Between these two values (for $c_s$ constant)
\begin{equation}
      v(z,t)=\tanh y=\frac{z/t+ c_s}{1+ c_sz/t}.
      \label{eq:v}
         \end{equation} 
and  
     \begin{equation}
\epsilon(z,t)=\epsilon_0 \left(e^{-c_sy}\right)^{\frac{1+c_s^2}{c_s^2}}.
\label{eq:eps}
     \end{equation} 
Equations \eqref{eq:v} and \eqref{eq:eps} take on some particular values that
are  useful to note for numerical checks
\begin{itemize}
\item  
For $z=-c_0\,t$:  $v=0=y$ and $\epsilon=\epsilon_0$.
\item For $z=t$:
$v= 1$, ($y\rightarrow \infty $) and
$\epsilon= 0$.
 \item  For
   $z=0$, whatever $t$: $v=c_s$ and $ \epsilon=\epsilon_0 \left(e^{-c_s\tanh^{-1} c_s}\right)^{\frac{1+c_s^2}{c_s^2}}$.
\end{itemize}


The relativistic SPH equations \cite{aguiar} are rather 
similar to the non-relativistic case, i.e. eq. \eqref{eq:sphrho} and \eqref{eq:eulersph} and read
\begin{equation}
s^*_i=\sum_{j=1}^N \nu_j W_{ij}
\label{eq:srel}
\end{equation}
and
\begin{equation}
  \frac {dP_i}{dt}=-\sum_{j=1}^N \nu_j\left[ \frac{p_i}{s^{*2}_i}+\frac{p_j}{s^{*2}_j}\right]\frac{d }{d z_i} W(z_i-z_j,h).
\label{eq:Prel}
  \end{equation}
Rather than mass density, entropy density is used as a conserved quantity. $s^*=\gamma s$ is the lab frame entropy density.
Rather than the time derivative of the velocity, the equation of motion involves a time derivative of $P=(\epsilon+p)\gamma^2v/s^*$, with $\epsilon$ the energy density and $p$ the pressure. In the following, we assume $c_s=1/\sqrt{3}$.

\begin{figure}[h!]
\centering
\includegraphics[width=0.46\linewidth]{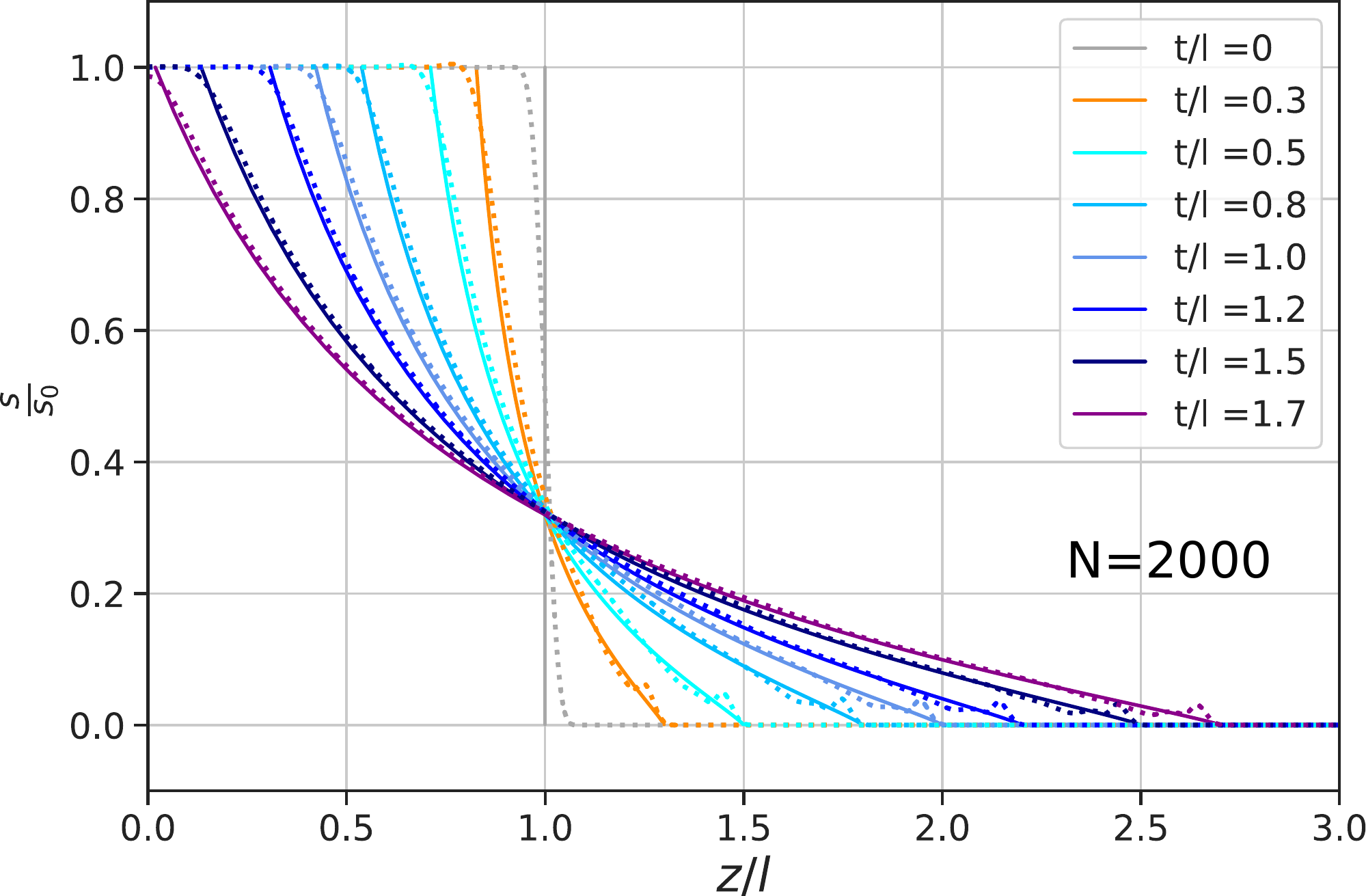}
  \includegraphics[width=0.46\linewidth]{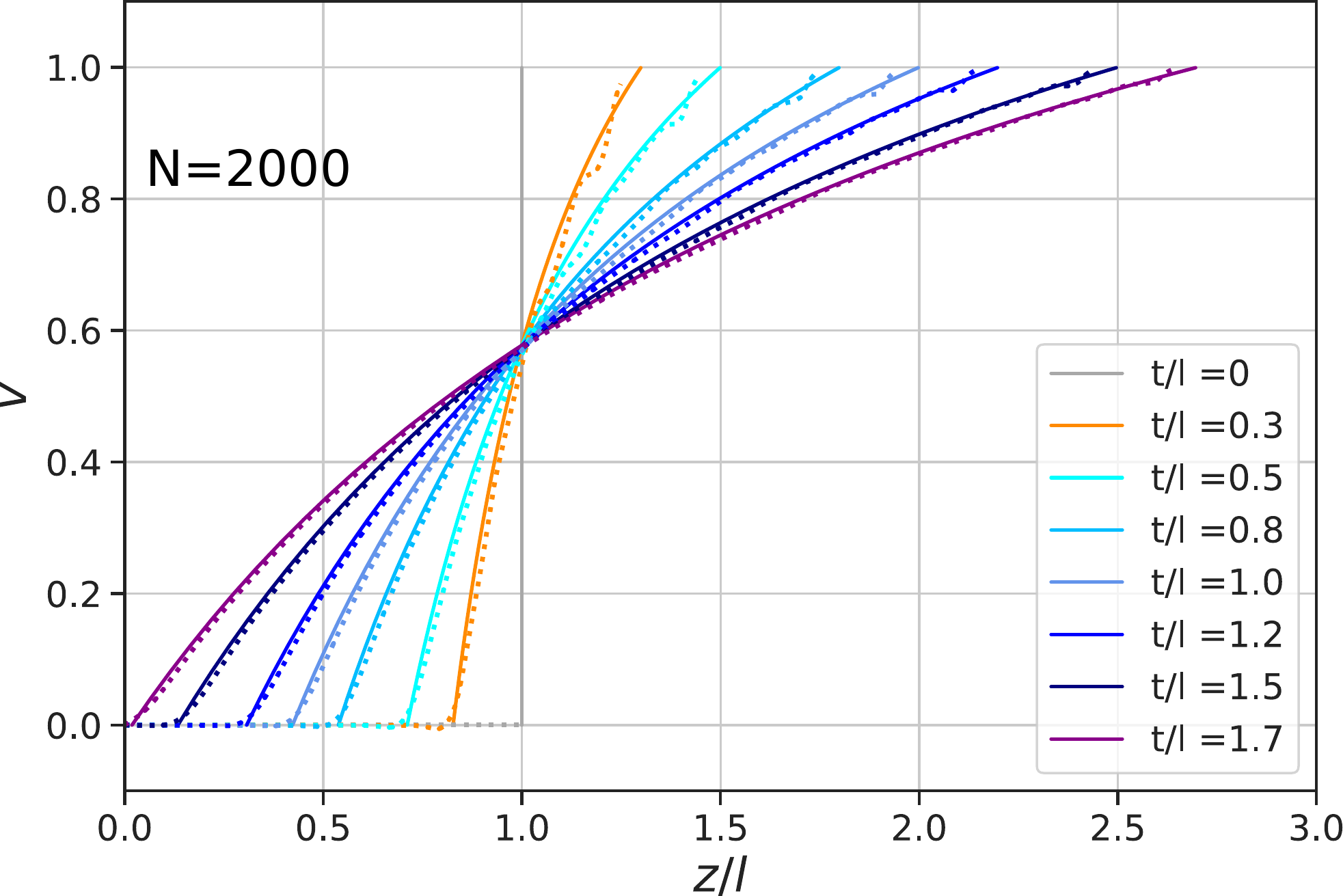}

\includegraphics[width=0.46\linewidth]{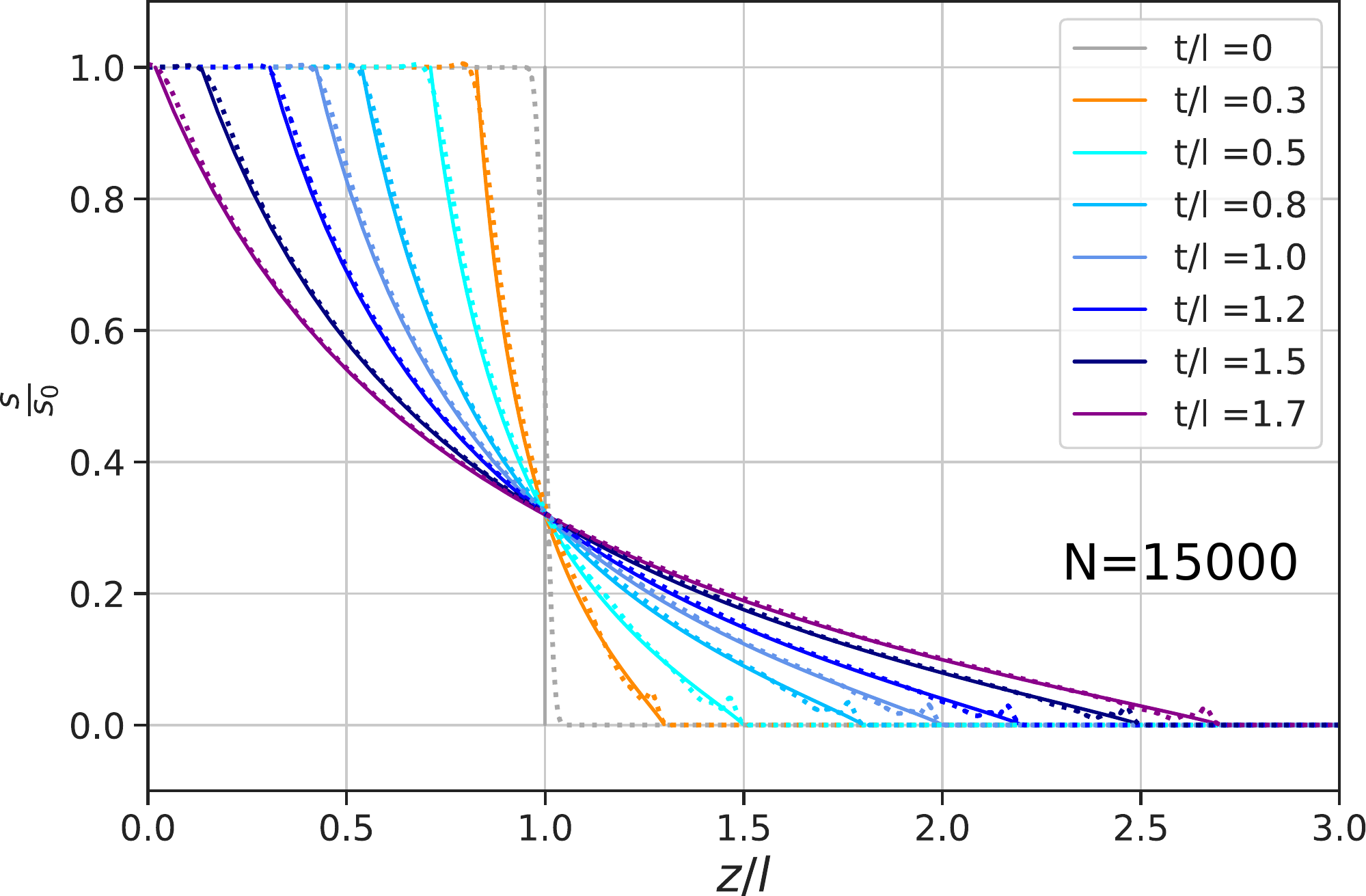}
  \includegraphics[width=0.46\linewidth]{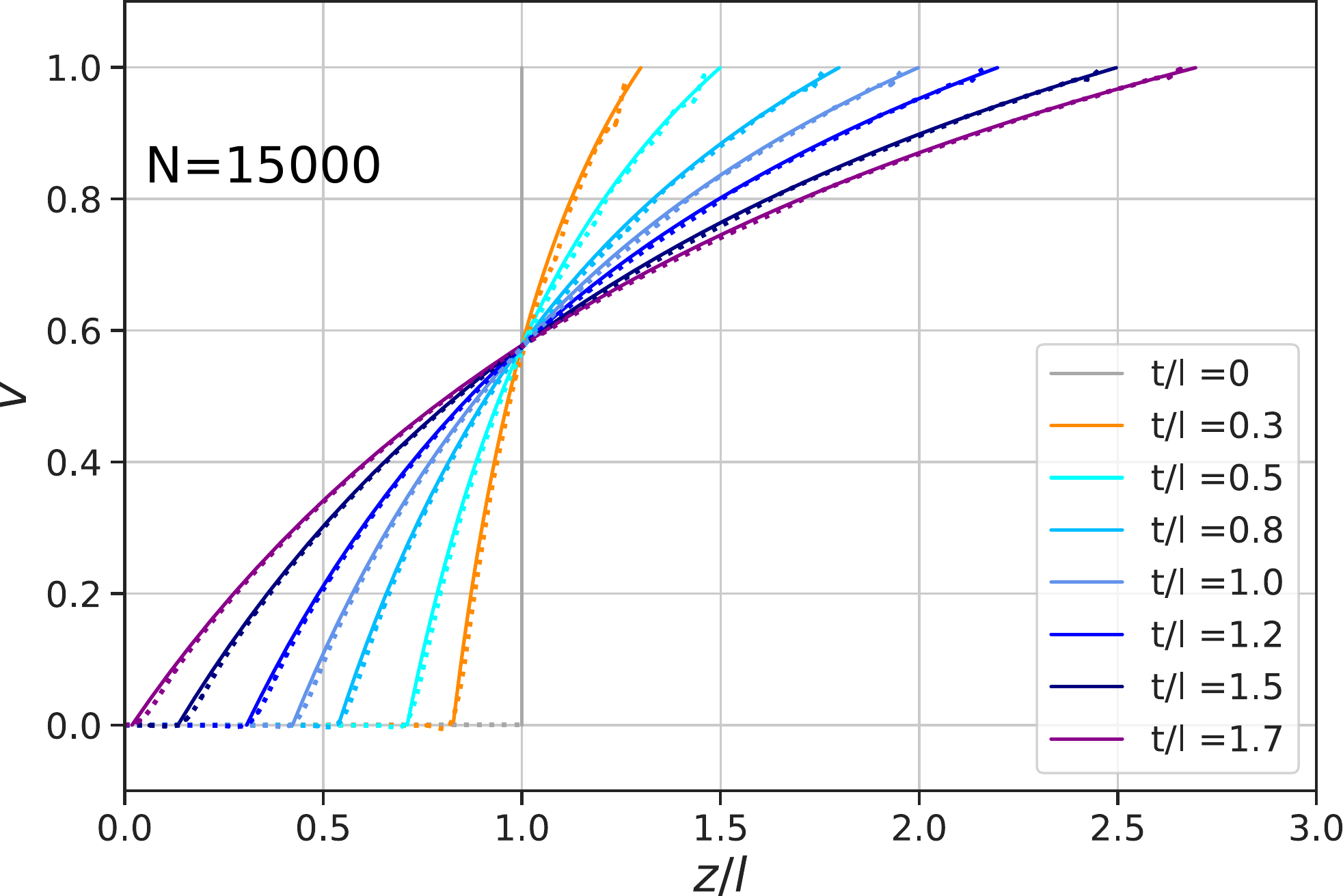}
\end{figure}
\begin{figure}[h!]
\centering
\includegraphics[width=0.46\linewidth]{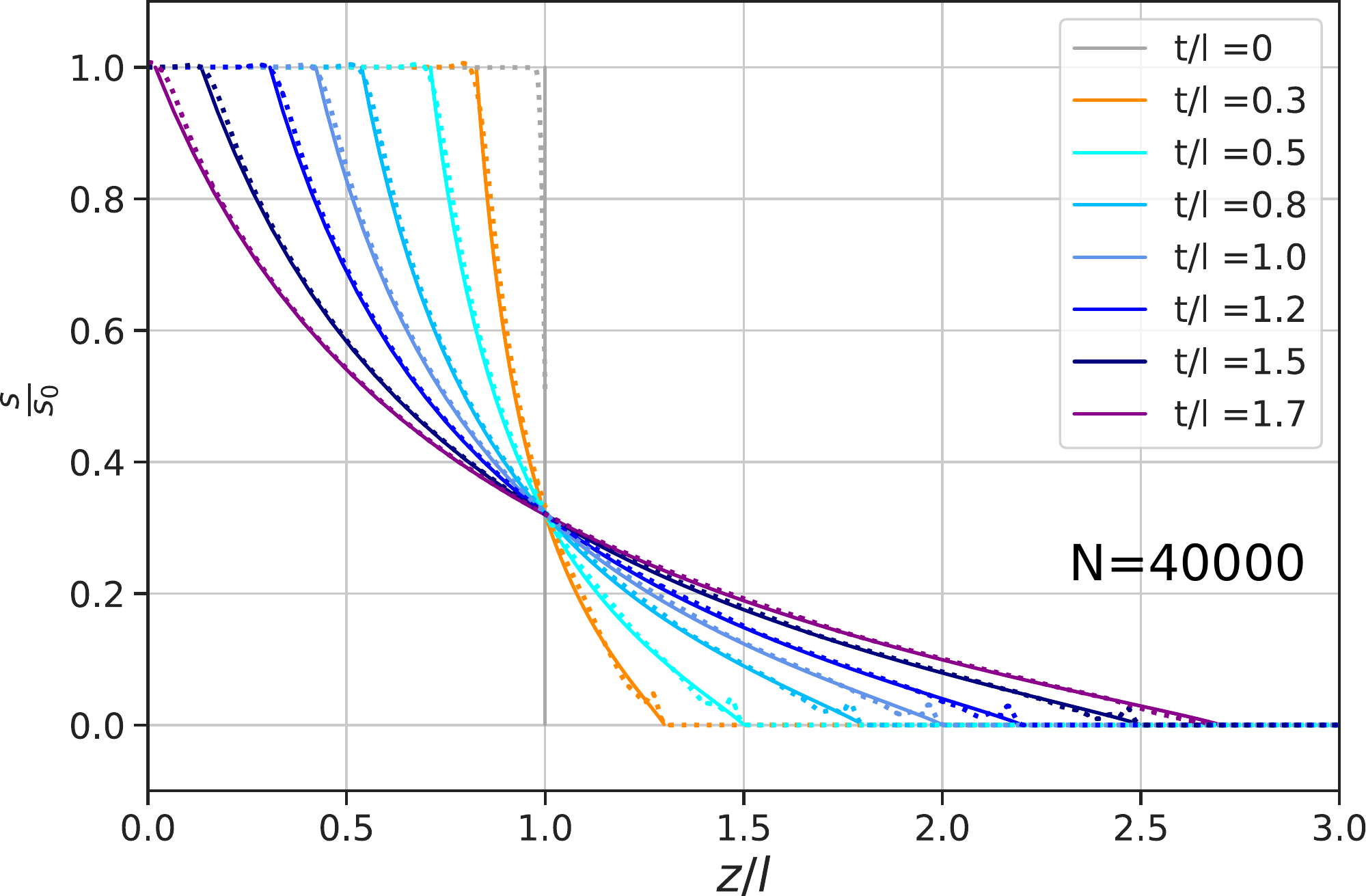}
  \includegraphics[width=0.46\linewidth]{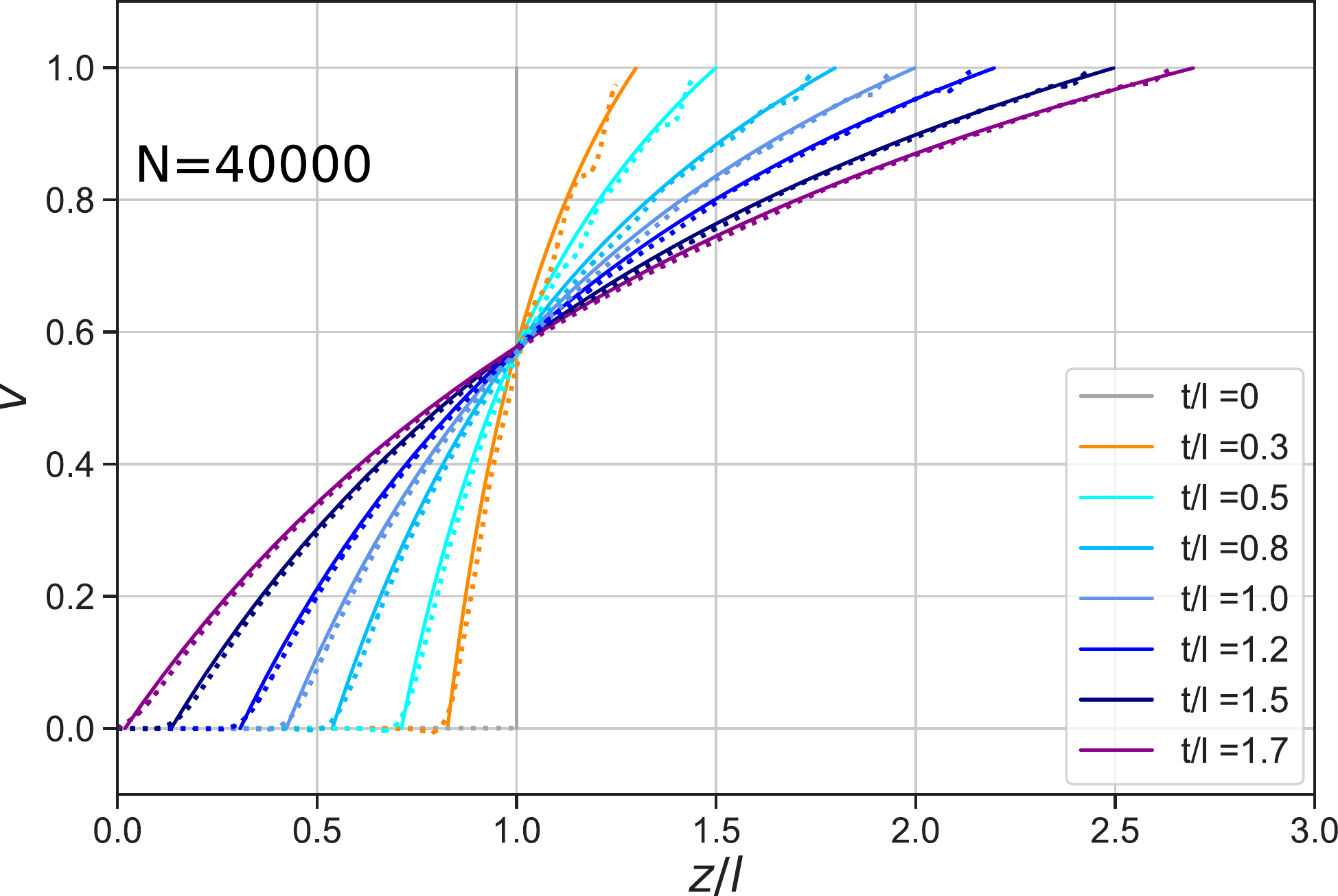}
  \caption{Solutions for (1+1)D relativistic gas expansion in vacuum: entropy density (left) and fluid velocity (right). Solid lines represent the exact solution, dotted lines display the result of SPH evolution with different number of interpolating points: $N=2000$ (upper), $N=15000$ (middle), $N=40000$ (lower).}
\label{fig:1dreltubec}
\end{figure}

Various solutions are shown in fig. \ref{fig:1dreltubec} with fixed smoothing length $h$ and three values of the number $N$ of SPH particles. Again large $N$ allows for a better solution but  is time-consuming in higher dimensions. In addition, increasing $N$ to 40 000 leads to little improvement. We note that the maximum value for $v/c_0$, namely $1$, is reached even for a modest value of $N$, contrarily to the classical case. Though these results are satisfying we are studying the relativistic implementation of a variable $h$ to check if we can get a precise and fast solution.

\section{Conclusion}
In this paper, we study technics to reproduce with the SPH method the exact solutions for a classical and a relativistic simple waves. These studies will be useful to benchmark the three-dimensional hydrodynamic code we are developing to simulate relativistic nuclear collisions. Increasing the number of fluid particles allows for a better solution but is time consuming. Variable $h$ solutions
allow for a better and faster solution but still need improvement.

\ack{Acknowledgments} 
This work is supported by Funda\c{c}\~ao de Amparo \`a Pesquisa do Estado de S\~ao Paulo (FAPESP) through  grants  2018/24720-6, 2020/08937-5, 2020/15893-4, project INCT-FNA Proc.~No.~464898/2014-5, University of S\~ao Paulo PUB grants 43-IF/910, 43-IF/913,
CNPq/PIBIC grant 2021/1548 and Helmholtz Forschungsakademie Hessen f\"ur FAIR.

\end{document}